\documentclass[preprintnumbers, prd, twocolumn, showpacs, floatfix, preprintnumbers, letterpaper, amsmath, amssymb, superscriptaddress]{revtex4}

\usepackage{amssymb, amsmath, bm, dcolumn, epsf, graphicx, latexsym, mathbbol, slashed, simplewick}

\usepackage{color}

\newcommand{\be}{\begin{equation}}
\newcommand{\ee}{\end{equation}}
\newcommand{\bea}{\begin{eqnarray}}
\newcommand{\eea}{\end{eqnarray}}
\newcommand{\barr}{\begin{array}}
\newcommand{\earr}{\end{array}}

\begin{document}

\title{ CMB Power Asymmetry from Primordial Sound Speed Parameter}

\author{Yi-Fu Cai}
\affiliation{Department of Physics, McGill University, Montr\'eal, QC H3A 2T8, Canada}

\author{Wen Zhao}
\affiliation{Key Laboratory for Researches in Galaxies and Cosmology, Department of Astronomy, University of Science and Technology of China, Hefei, Anhui, 230026, China}

\author{Yang Zhang}
\affiliation{Key Laboratory for Researches in Galaxies and Cosmology, Department of Astronomy, University of Science and Technology of China, Hefei, Anhui, 230026, China}


\begin{abstract}
The hemispherical power asymmetry in the cosmic microwave background may be explained by the modulation of some primordial cosmological parameter, such as the sound speed of the inflaton. This modulation can be achieved by the perturbation of a light field during inflation. 
We numerically examine this mechanism and show it can be consistent with current observations. Further, this model predicts that the power asymmetry also exists in the temperature-polarization correlation and polarization autocorrelation with the same shape, and in primordial non-gaussianity of equilateral type with a particular shape. Therefore, our mechanism is observationally detectable in forthcoming experiments.
\end{abstract}

\maketitle


The recent released Planck data reported a hemispherical power asymmetry in the cosmic microwave background (CMB) fluctuations \cite{Ade:2013sta}, and provided a much stronger evidence on this anomaly than it was earlier reported in the WMAP data. Such a power asymmetry can be modeled as a dipolar modulation of a statistically isotropic CMB sky in terms of temperature fluctuations in direction $\hat{n}$ \cite{Ade:2013sta}:
\begin{eqnarray}\label{T_kr}
 \frac{\Delta T}{T}(\hat{n}) = s(\hat{n}) \left[ 1 + A ~\hat{n} \cdot \hat{p} \right] ~,
\end{eqnarray}
where $s(\hat{n})$ is a statistically isotropic map, $A$ characterizes the amplitude of dipolar asymmetry, and $\hat{p}$ is its direction. To translate to the expression of the primordial power spectrum, the modulation required to explain this asymmetry can be written as a spatially-varying power spectrum \cite{Dai:2013kfa},
\begin{eqnarray}\label{P_kr}
 P(k, \vec{r}) = P(k) \left[ 1 + 2A ~\hat{p} \cdot {\vec{r}} / {r_{ls}} \right]~,
\end{eqnarray}
where $r_{ls}$ is the distance to the last scattering surface.

The best-fit dipolar asymmetry has an anisotropy direction $(227,-27)$ and the corresponding amplitude is given by $A=0.072 \pm 0.022$ for the CMB power with $l \lesssim 64$ (and thus $k \lesssim 0.035 {\rm Mpc}^{-1}$) \cite{Ade:2013sta}. However, the asymmetry does not necessarily exist at smaller length scales. Particularly, the constraint from the Sloan Digital Sky Survey quasar sample \cite{Hirata:2009ar} requires $A<0.0153$ ($99\%$ C.L.) for the power asymmetry oriented in the direction of the CMB dipole in which the typical wavenumber is $k \sim 1 {\rm Mpc}^{-1}$. Thus, any model that accounts for the CMB power asymmetry has to produce a strong scale dependence so that it can be in agreement with both the CMB and the quasar constraints.

As pointed out in \cite{Erickcek:2008sm}, a single-field slow-roll inflation model cannot generate such an asymmetry without violating the constraints to the homogeneity of the Universe. The same paper also proposed a so-called Erickcek-Kamionkowski-Carroll (EKC) mechanism based on a curvaton model \cite{Lyth:2001nq, Mollerach:1989hu} and thus can explain this anomaly without violating the homogeneity constraint. However, the original model is inconsistent with the quasar bound since the signature is scale-independent. Also, the model leads to a large value of the non-Gaussianity parameter which has been ruled out by Planck \cite{Ade:2013ydc}. Instead, an improved curvaton model in which the curvaton decay takes place after dark matter freezes out was studied \cite{Erickcek:2009at}. In the presence of super-Hubble iso-curvature fluctuations, the power asymmetry can be obtained because of the difference between the value of curvaton field on one side of the last scattering surface and its average value in the observable Universe. Accompanied with this anomaly, the model predicts an iso-curvature contribution to primordial perturbations that may need a fine-tuning on the model's parameters in order to be consistent with the Planck data. Moreover, it is shown in Ref. \cite{Schmidt:2012ky} that the non-Gaussianities of primordial perturbations may give rise to such a power asymmetry if the squeezed limit of the bispectrum is sufficiently divergent.

It was exquisitely observed in \cite{Dai:2013kfa} that the power asymmetry may arise from a modulation of any cosmological parameters that affect the CMB power spectrum, and a comprehensive analysis was performed that includes iso-curvature perturbations, gravitational waves, a variation of spectral index, a dipolar modulation of the re-ionization optical depth, and a compensated baryon density. If the value of one cosmological parameter on one side of the CMB sky is different from the one on the other side and the power spectrum is correlated with this parameter, then the power spectrum in one side may differ from that on the other side as well. 
%
%

Despite of the above models, we suggest that the power asymmetry may be explained by a modulation of the sound speed parameter $c_s$ of primordial inflationary perturbations. In general, an inflation model can be realized by a K-essence field minimally coupled to Einstein gravity \cite{ArmendarizPicon:1999rj}, with the Lagrangian in form of $P(X, \phi)$ and $X\equiv -g^{\mu\nu}\partial_\mu\phi\partial_\nu\phi/2$. This model may come from some stringy motivation such as an effective description of D-brane dynamics \cite{Aharony:1999ti}, or from the effective single-field description of a coupled multi-field system where the heavy modes are integrated out \cite{Tolley:2009fg}. From time being, let us put aside its theoretical origin and focus on the phenomenological implication on CMB.

For this type of models, the gradient stability of the inflaton fluctuation is characterized by the sound speed, the square of which is defined as
 $c_s^2 \equiv {P_{,X}} / (P_{,X}+2XP_{,XX})$.
The subscript ``$_{,X}$" denotes the derivative with respect to $X$. To deal with such an inflationary dynamics, it is convenient to introduce the following slow-roll parameters:
 $\epsilon = -{\dot{H}}/{H^2} $, $\eta =  {\dot \epsilon}/{H\epsilon}$, $s = {\dot c_s}/{Hc_s}$,
where $H$ is the Hubble parameter defined as $\dot a/a$. The amplitude of the field fluctuation during inflation is determined by the Hubble rate through $\delta\phi \simeq H/2\pi$, but the freeze-out moment depends on when the perturbation mode exit the sound horizon $k^{-1} \simeq c_s/H$. Therefore, the amplitude of primordial power spectrum is given by
\begin{eqnarray}\label{P_zeta}
 P_\zeta = \frac{P_{\zeta,0}}{c_s}~,~~P_{\zeta,0} = \frac{H^2}{8\pi^2M_p^2\epsilon}~,
\end{eqnarray}
where we introduce $P_{\zeta,0}$ which has identical form as the power spectrum of a canonical inflation model. Intuitively, a power asymmetry can be obtained if the value of $c_s$ on one side of the CMB sky differs from the one on the other side. In order to be consistent with the parametrization of the polar asymmetry as introduced in Eq. \eqref{P_kr}, we need a spatially varying sound speed as follows,
\begin{eqnarray}\label{c_s_A}
 c_s(k,\vec{r}) = \bar{c}_s(k) \left[ 1 + 2A ~\hat{p} \cdot {\vec{r}} / {r_{ls}} \right]^{-1} ~,
\end{eqnarray}
where $\bar{c}_s(k)$ is the direction independent part of the sound speed for inflationary perturbations in our observed Universe, and in general it depends on the wavenumber $k$. 

Now we use the model of {\it multi-speed inflation} \cite{Cai:2009hw} to illustrate the possibility of this sound speed modulation. We phenomenologically consider a double-field inflation model, with one field being described by a DBI action and the other by a canonical field. The total action is constructed by the sum of the two. Therefore, we take the Lagrangian density of two fields as,
\begin{small}
\begin{eqnarray}
 {\cal L} = \frac{1}{f(\phi, \chi)} (1-\sqrt{1+f\partial_\mu\phi\partial^\mu\phi}) - \frac{1}{2}\partial_\mu\chi\partial^\mu\chi - V(\phi, \chi)~.
\end{eqnarray}
\end{small}
Here, $\phi$ plays the role of the inflaton field and $\chi$ is an entropy field which does not contribute to the background evolution. The DBI-type kinetic term for $\phi$ involves a coefficient $f$. This coefficient is often interpreted as a warping factor from the point of view of string cosmology, but right now let us assume it as a function of $\phi$ and $\chi$ for phenomenological consideration.
A first interesting property is that these two fields carry different values of sound speed parameters. For $\chi$, there is $c_s^\chi = 1$ since its kinetic term is canonical; however, for the inflaton, the sound speed takes the form of,
\begin{eqnarray}\label{cs^phi}
 c_s(t) = \sqrt{1-f(\phi, \chi) \dot\phi^2}~.
\end{eqnarray}
This model was extensively studied in \cite{Pi:2011tv} and was expected to produce a large value of primordial equilateral non-gaussianity \cite{Emery:2013yua} due to the following relation\footnote{We adopt the convention of primordial non-gaussianity from \cite{Ade:2013ydc}.},
\begin{eqnarray}\label{f_NL}
 f_{\rm NL}^{\rm DBI} = - \frac{35}{108} (\frac{1}{c_s^2}-1)~.
\end{eqnarray}
Unfortunately, the interest in such a model was evaded after Planck since no evidence was found to prove the existence of primordial non-gaussianities. However, we show that this model may be applied to explain the hemispherical asymmetry if the value of sound speed varies from one side of the sky to the other.

From \eqref{cs^phi}, the sound speed can be modulated by the field fluctuation $\Delta\chi$. Especially, due to the Grishchuk-Zel'dovich (GZ) effect\cite{Grishchuk:1978}, the $\Delta\chi$ modes at very large scales could bring an enhancement within the observable universe which is expected as an approximately linear function of position. In \cite{Lyth:2013vha}, this effect was considered to explain the dipolar anomaly through a curvaton mechanism where the primordial fluctuation of $\chi$ has to be responsible for curvature perturbation as well. Thus, the enhancement factor has to be finely tuned to generate the required asymmetry while the CMB quadrupole is still small enough to accommodate with observation. In our case, it is not necessary that the field fluctuation $\delta\chi$ be responsible for the curvature perturbation. Thus we do not need to require a manifest enhancement on very large scales. Instead, the asymmetry can arise from the so-called warping factor $f(\phi, \chi)$.

By expanding \eqref{cs^phi} to linear order, one easily derives:
\begin{eqnarray}\label{c_s_linear}
 c_s 
  = \bar c_s(t(k)) \left[ 1+ \frac{54}{35} f_{\rm NL}^{\rm DBI} \frac{f_{,\chi}}{f} \Delta\chi \right] ~,
\end{eqnarray}
where the relation \eqref{f_NL} was applied. Note that $\bar c_s$ slowly varies as a function of the cosmic time during inflation and thus becomes $k$ dependent for  perturbation modes at Hubble-exit. According to the GZ effect, one expects the field fluctuation to be\cite{Lyth:2013vha}:
\begin{eqnarray}\label{deltachi_GZ}
 \Delta \chi \simeq E P_{\delta\chi}^{\frac{1}{2}} ~\hat{p} \cdot {\vec{r}} / {r_{ls}} ~,
\end{eqnarray}
at very large scales. During inflation, there is always an approximate relation $P_{\delta\chi}^{\frac{1}{2}} = {H}/{2\pi}$ at the moment of Hubble-exit. The coefficient $E$ is viewed as an enhancement factor of GZ effect, and is found to be tightly constrained by observations if both the asymmetry and power spectrum are generated by the same field as has been analyzed in \cite{Lyth:2013vha}. There is, however, no excuse to require the existence of a very large value of $E$. Thus, in the following we simply take $E \sim O(1)$ which easily satisfies the bound provided in \cite{Lyth:2013vha}.

Inserting the field fluctuation \eqref{deltachi_GZ} into the linearized expansion of the sound speed \eqref{c_s_linear} and comparing with the expected form \eqref{c_s_A}, we get,
\begin{eqnarray}
 |A(k)| \simeq \frac{27}{70\pi} E(k) H(k) |f_{\rm NL}^{\rm DBI}(k)| |\frac{f_{,\chi}}{f}(t_k)|~,
\end{eqnarray}
where the amplitude of the hemispheric asymmetry is a function of the comoving wavenumber. Namely, the coefficients $E$, $H$, and $f_{\rm NL}^{\rm DBI}$ at Hubble-exit can be $k$ dependent, although such a dependence is negligible in usual inflation models. Moreover, the $k$ dependence also comes from the evolution of the $\chi$ field during inflation, because when the primordial perturbation modes exit the Hubble radius at different time the corresponding value of the $\chi$ field is different. This remarkable feature provides a physical explanation for the observational fact that the power asymmetry is only significant at cosmological scale but becomes small at the ${\rm Mpc}$ scale.

Specifically, we consider an example of power law function, i.e., $f(\chi) \sim \chi^p$. Then, using \eqref{P_zeta} and \eqref{f_NL} we obtain
\begin{eqnarray}\label{A_bound}
 |A(k)| \lesssim \frac{27 \sqrt{|1-n_s|}}{35\sqrt{2}} 
 \frac{ E P_\zeta^{\frac{1}{2}} |f_{\rm NL}^{\rm DBI}|}{ ( 1 - \frac{108}{35}f_{\rm NL}^{\rm DBI})^{\frac{1}{4}}} 
 \frac{p M_p}{\chi(t_k)}~,
\end{eqnarray}
where the bound on the slow roll parameter from the spectral tilt has been applied. According to the Planck data \cite{Ade:2013ydc, Ade:2013rta}, we learn that $n_s = 0.9603 \pm 0.0073$ at $1\sigma$ and $\bar c_s \geq 0.07$ at $2\sigma$ and the best fit value of the power spectrum gives $P_\zeta = 22.1536 \times 10^{-10}$. Substituting these values into the inequality yields an upper bound,
 $|A(k)| \lesssim 8.908 \times 10^{-5} {\chi(t_k)}^{-1} ~pE ~M_p$~.
Typically, we have $p\sim O(1)$. Moreover, the constraint on the CMB quadrupole does not favor a manifest enhancement on the amplitude of field fluctuation at very large scale and thus we may typically assume $E\sim O(1)$ as well. Eventually, we can use the value of the $\chi$ field to generate the required power asymmetry. Namely, for the perturbation mode with $k_a \sim 0.035 {\rm Mpc}^{-1}$ exiting the Hubble radius, we expect,
\begin{eqnarray}
 \chi({t_{k_a}}) \gtrsim 0.00124 ~pE~M_p~;
\end{eqnarray}
when the perturbation mode with $k_b \sim 1 {\rm Mpc}^{-1}$ exit the Hubble radius, we then need,
\begin{eqnarray}
 \chi({t_{k_b}}) \gtrsim 0.00582 ~pE~M_p~.
\end{eqnarray}
Then, the desired asymmetry can accommodate both the CMB and the quasar observations. Such a result suggests that the vacuum of the $\chi$ field has to be away from the origin and $\chi$ evolves from a small value to a large one during inflation. Phenomenologically, this is easy to be achieved such as in a small field model.


{\it Numerical estimate:}
\begin{figure}
\begin{center}
\includegraphics[width=7.5cm]{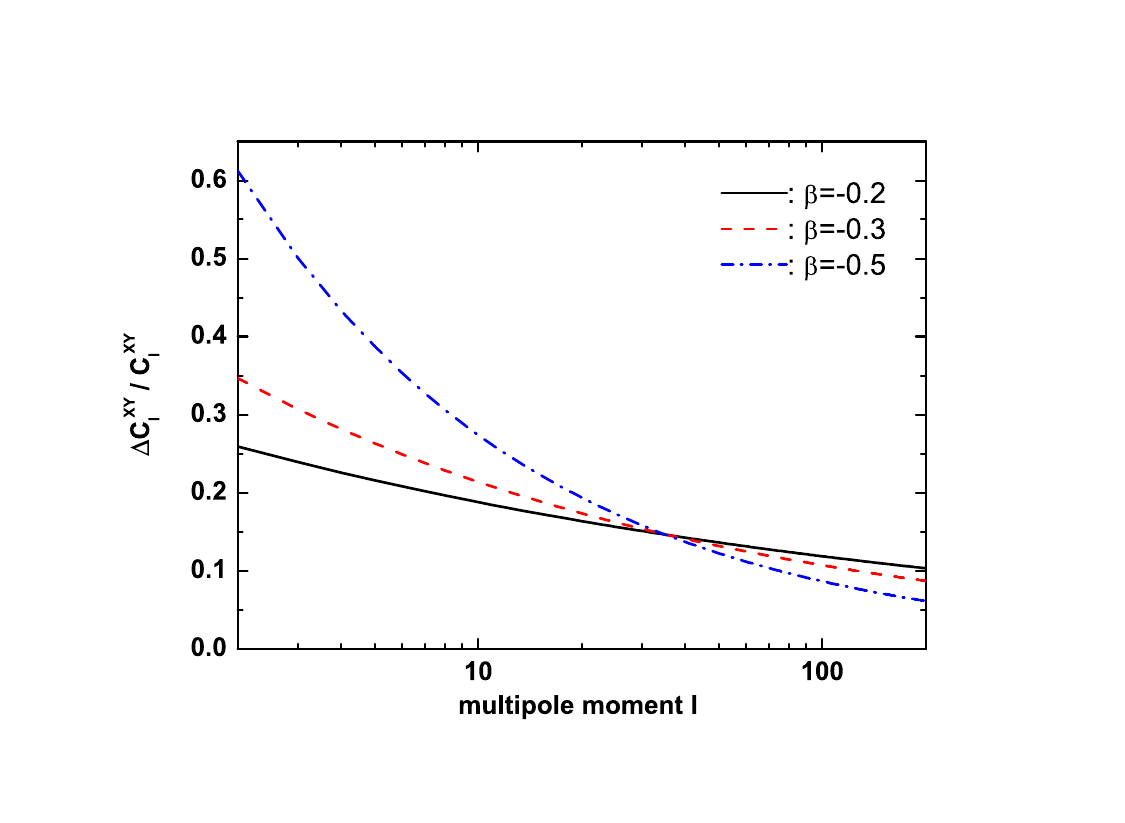}
\includegraphics[width=7.5cm]{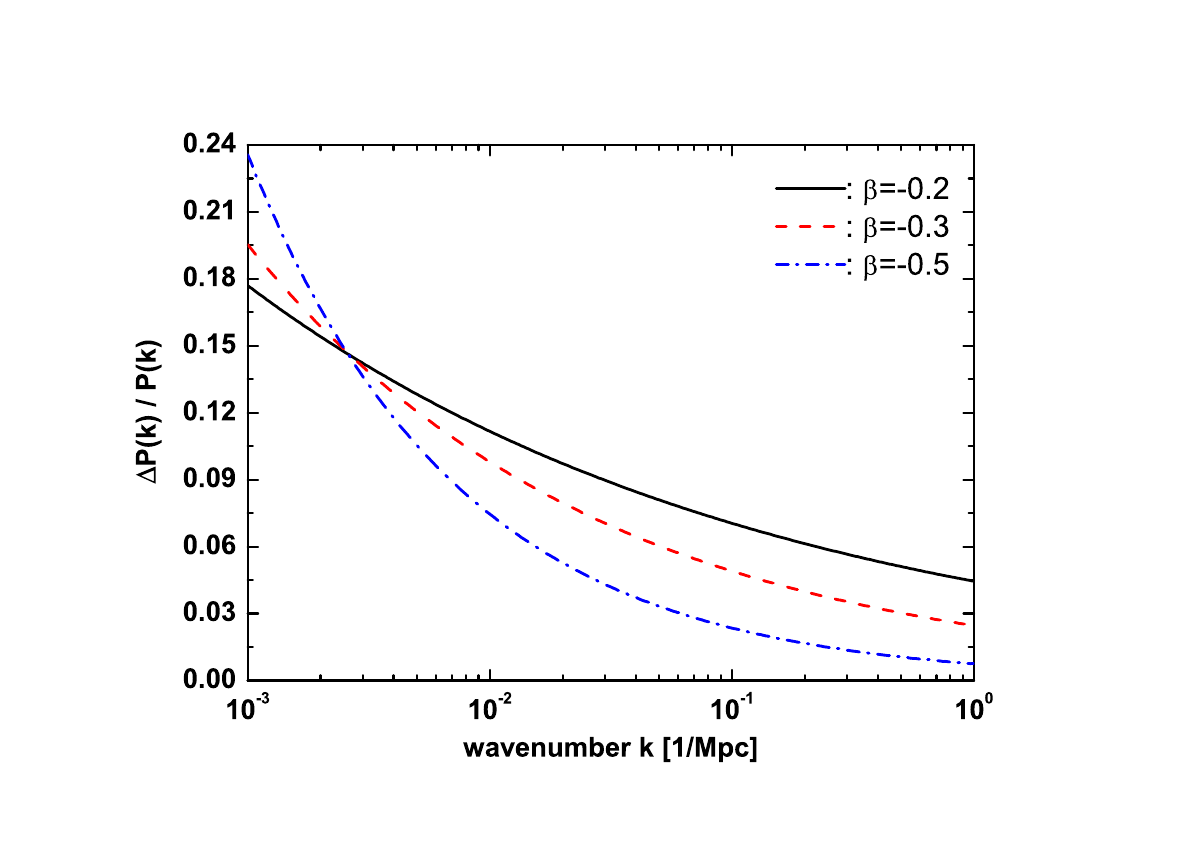}
\caption{\label{fig1} The fractional changes $\Delta C_{l}^{\rm XY}/C_{l}^{\rm XY}$ (XY=TT,TE,EE) in the CMB power spectra (upper) and the fractional change $\Delta P(k)/P(k)$ in the matter power spectrum (lower) for models with different $\beta$ in our mechanism. Each curve is normalized so that $A=0.072$.}
\end{center}
\end{figure}
Following the previous model, the power asymmetry may arise from a modulation of the sound speed that affects the CMB power spectrum without modifying the inflationary background. Therefore, one expects that the current constraint on inflationary models can be safely satisfied. However, since our model allows the value of $c_s$ on one side of the sky to be different from that on the other side, the corresponding CMB power spectrum can be different on the two sides.
Specifically, we would like to map the modulation of $c_s$ into the $\Delta C_l^{\rm TT}$ through the relation
 $\Delta C_l^{\rm TT} = \frac{ \partial C_l^{\rm TT} }{\partial c_s} \Delta c_s$.
%
The CMB power spectrum is calculated by $C_l^{\rm TT}=\int P_\zeta(k) \Delta_l^{\rm T2}(k) \frac{dk}{k}$, where $\Delta_l^{\rm T}(k)$ is the transfer function, and is independent of the primordial power spectrum $P_\zeta(k)$. We assume that the sound speed can be written in form of $c_s(k,{\vec r})=\tilde{c}_s(1+D(k/k_0,{\vec r}))$, where $\tilde{c}_s$ is constant, $D(k/k_0,{\vec r})$ is the small correction term, and $k_0=0.74\times 10^{-4}$Mpc$^{-1}$ is the chosen pivot wavenumber. Employing the relation in Eq.(\ref{P_zeta}), we arrive at
 $\Delta C_l^{\rm TT}/C_{l}^{\rm TT} \simeq -D(l,{\vec r})$,
where we have used that the transfer function is sharply peaked at value $l\simeq k/k_0$, which is a reflection of the fact that metric fluctuations at a particular linear scale $k^{-1}$ lead to CMB anisotropies predominantly at angular scales $\theta \sim kd$ (where $d$ is the distance to the surface of last scattering). 

In order to estimate the effect, we ignore the dependence on ${\vec r}$ and assume a power-law form for the function $D(x)=\alpha x^{\beta}$, where the index $\beta<0$, since we expect that the anisotropy effect is obvious only at the large scales. Following \cite{Dai:2013kfa}, the amplitude parameter $\alpha$ can be fixed by the asymmetry parameter $A=0.072$, which is determined from the data weighting in all spherical harmonic modes equally up to $l_{\rm max}=64$, i.e.,
 $A = \frac{1}{2 N } \sum_{l=2}^{l_{\rm max}} (2l+1) \frac{ \Delta C_l^{\rm TT} }{ C_l^{\rm TT} }~,$
with $N = \sum_{l=2}^{l_{\rm max}} (2l+1)$.
In the upper panel of Fig. \ref{fig1}, we show the fractional power-spectrum differences, where the cases with $\beta = -0.2, ~-0.3, ~-0.5$ are considered. The corresponding fractional change $\Delta P(k)$ in the matter power spectrum induced by the modulations in different cases are presented in the lower panel of the same figure. As expected, we find that these modified the CMB power and matter power spectra only on small scales.

{\it Prediction:}
In addition to the temperature autocorrelation, there are also the temperature-polarization correlations ($C_{l}^{\rm TE}$) and polarization autocorrelation ($C_{l}^{\rm EE}$). By a similar analysis as done above, we find that in this model, the fractional changes $\Delta C_{l}^{\rm TE}/C_{l}^{\rm TE}$ and $\Delta C_{l}^{\rm EE}/C_{l}^{\rm EE}$ should be exactly the same as those of $\Delta C_{l}^{\rm EE}/C_{l}^{\rm EE}$, which provide an excellent opportunity to test this model with the polarization observations.

Another interesting prediction of this model is related to the primordial non-Gaussianity parameter of equilateral type, which takes $ f_{\rm NL}^{eq} \simeq \frac{1}{3} (c_s^{-2}-1)$. It was expected that a large value of primordial non-gaussianities may be generated by this model if one tunes $c_s$ to be a small quantity such as in the model of DBI inflation, and thus the corresponding models now is strongly constrained by the Planck data. However, if we expect that there exists a modulation of sound speed which accounts for the power asymmetry, then we also reach an interesting conclusion that the power of bispectrum is asymmetric as well. In particular, in this model the value of $f_{\rm NL}^{eq}$ is scale-dependent, and also in the large scales, the asymmetry of $f_{\rm NL}^{eq}$ should be significant. This signature, together with the signature of polarization modes, can be tested by future observations.


{\it Conclusion:}
Until now, it is still unknown whether a hemispherical asymmetry in the CMB fluctuations is associated with the background of observational data or indicates nontrivial physics beyond the standard scenario. While keeping aware of the foreground contamination of data, it deserves to question plausible physical explanations for generating such an asymmetry. After Planck Collaboration reported this anomaly in the data release, a few papers appeared which attempted to provide a physical mechanism, namely see \cite{Dai:2013kfa, Lyth:2013vha, Notari:2013iva, Wang:2013lda, Liu:2013kea, McDonald:2013aca, Namjoo:2013fka, Chen:2013eaa, Liddle:2013czu, D'Amico:2013iaa}.

In this Letter we explore another mechanism of generating the hemispherical asymmetry in the CMB fluctuations by requiring a statistically inhomogeneous sound speed parameter. 
To illustrate the feasibility of our mechanism, we phenomenologically consider a model of {\it multi speed inflation} which involves two scalar fields. One of them is assumed not to contribute to the inflationary background at all. However, it couples to the kinetic term of the inflaton field and its fluctuation can modulate the inflaton's sound speed. After that, the EKC mechanism can easily yield a potentially statistical anisotropy on the power spectrum. By performing a numerical estimate, we show that the model can generate the power asymmetry in agreement with both the CMB observation and the quasar constraint. Furthermore, the model gives promising predictions on the polarization correlations and bispectrum and thus observations of CMB polarization and the large-scale structure will be significant on improving constraints on this anomaly.

{\it Acknowledgments:}
We are grateful to R. Brandenberger, P. Chen, J. Emery, E. Ferreira, G. Tasinato and D. Wands for useful discussions. Y.F.C. is supported in part by the Department of Physics in McGill University. W.Z. is supported by NSFC No.11173021, 11075141 and project of KIP of CAS. Y.Z. is supported by NSFC No. 10773009, SRFDP, and CAS.

\end{document}